\documentclass[aps,pra,twocolumn,floats,showpacs,superscriptaddress,10pt]
{revtex4-1}

\usepackage{bm}
\usepackage{graphicx}
\usepackage{amssymb}
\usepackage{amsfonts}
\usepackage{amsmath}
\usepackage{epstopdf}
\usepackage{color}

\def \IFPAN{Institute of Physics, Polish Academy of Sciences, al. 
Lotnik\'{o}w 32/46, 02-668 Warsaw, Poland}

\begin{document}

\title{Calculated optical properties of Co in ZnO: internal and 
ionization transitions}

\author{A. Ciechan}\email{ciechan@ifpan.edu.pl}
\author{P. Bogus{\l}awski}\email{bogus@ifpan.edu.pl}\affiliation{\IFPAN}
 
\date{\today}

\begin{abstract} 
Previous luminescence and absorption experiments in Co-doped ZnO revealed two ionization and one intrashell transition of $d(\textrm{Co}^{2+})$ electrons. Those optical properties are analyzed within the generalized gradient approximation to the density functional theory. The two ionization channels involve electron excitations from the two $\textrm{Co}^{2+}$ gap states, the $t_{2\uparrow}$ triplet and the $e_{2\downarrow}$ doublet, to the conduction band. The third possible ionization channel, in which an electron is excited from the valence band to the $\textrm{Co}^{2+}$ level, requires energy in excess of 4~eV, and cannot lead to absorption below the ZnO band gap, contrary to earlier suggestions. We also consider two recombination channels, the direct recombination and a two-step process, in which a photoelectron is captured by $\textrm{Co}^{3+}$ and then recombines via the internal transition. Finally, the observed increase the band gap with the Co concentration is well reproduced by theory.  

The accurate description of ZnO:Co is achieved after including $+U$ corrections to the relevant orbitals of Zn, O, and Co. The $+U(\textrm{Co})$ value was calculated by the linear response approach, and independently was obtained by fitting the calculated transition energies to the optical data. The respective values, 3.4 and 3.0~eV, agree well.  Ionization of Co induces large energy shifts of the gap levels, driven by the varying Coulomb coupling between the $d(\textrm{Co})$ electrons, and by large lattice relaxations around Co ions. In turn, over $\sim 1$~eV changes of $\textrm{Co}^{2+}$ levels induced by the internal transition are mainly caused by the occupation-dependent $U(\textrm{Co})$ corrections.
\end{abstract}


\maketitle

\section{\label{sec1}Introduction}
ZnO doped with Co is studied since five decades. The detailed 
experimental investigations conducted by Koidl~\cite{Koidl} showed that 
Co in ZnO substitutes for Zn and acquires the $\textrm{Co}^{2+}$ ($d^7$) 
electronic configuration with spin 3/2.  Information about the Co-induced 
levels was provided by optical measurements. Both internal 
$d(\textrm{Co})$ and ionization transitions were observed. 
The intrashell line at about 2.0~eV originates in the 
$e_{2\downarrow}\to t_{2\downarrow}$ (i.e., $^4A_2\to ^4T_1$)  
transition, and typical is 
fine split~\cite{Koidl, Tuan2004, KIM2004, Pacuski:PRB2006, Singh2008, 
Gilliland, Schulz, Jin, Matsui2013, Guo2015, Xu2013, Lecuna2014}. The 
observed splittings, of the order of few tens of eV, are due to a 
combined effect of the crystal field and the weak spin-orbit coupling, 
and they are not always experimentally resolved. Next, there are two 
ionization transitions beginning just below the ZnO band 
gap~\cite{Tuan2004, Gilliland, Jin, Guo2015, Kittilstved}, which are also 
reflected in photoconductivity. 
Those sub-band gap optical transitions can lead to applications in 
photocatalysis and photovoltaics~\cite{Kumar2015, Samadi2016}. The 
internal excitations of $\textrm{Co}^{2+}$ are utilized in efficient 
hydrogen production by photoelectrochemical 
water-splitting~\cite{Jaramillo}. The near UV-visible photodetectors are 
fabricated with Co-doped ZnO nanoparticles~\cite{Salman2013, Jacob2017}.

Investigations of ZnO:Co were intensified by the discovery of 
ferromagnetism (FM) at room temperatures~\cite{Tseng, Lu2009, Ciatto, 
LiLi2012}.  
Mechanism of magnetic coupling depends critically on the 
sample microscopic morphology~\cite{Dietl, Sawicki}, and in particular on 
the presence of defects~\cite{Tseng, Lu2009, Ciatto, LiLi2012, Qi2011}. 
While the origin of FM in ZnO:Co is out of the scope of this paper, it is 
obvious that the electronic structure of Co determines both its charge 
and spin states as a function of the Fermi energy, thus providing a 
necessary basis for understanding magnetic coupling between Co ions.

Accurate and efficient theoretical description of transition metal (TM) 
dopants by first principles methods remains a challenging problem. The 
underestimation of the single-particle band gap $E_{gap}$ by the local 
density (LDA) and the generalized gradient (GGA) approximations in the 
density functional theory distorts the levels, accessible charge states, 
and ionization energies of TM ions in ZnO. This is the case of electronic 
structure of Co calculated in Refs~\cite{spaldin, walsh, Gilliland2012}. 
An efficient procedure improving the LDA or GGA electronic 
structure of the host as well as the properties of TM dopants consists is 
adding the $+U$ terms~\cite{Cococcioni}. 
Those terms can be treated 
as adjustable parameters, or can be calculated 
self-consistently~\cite{Cococcioni}, but they should be applied to all 
orbitals relevant for the problem. 
For example, in the case of the LDA$+U$ and GGA$+U$ 
calculations in which only the $d$ states of Zn ions are 
corrected~\cite{Gopal, Chanier, Iusan, sarsari, Gluba}, the band gap 
problem persists, 
and the predicted Co levels are not reliable. 
Two correction schemes giving a correct band gap of ZnO are the 
nonlocal external potential (NLEP) corrections~\cite{Lany, Raebiger} 
and the self-interaction-corrections (SIC)~\cite{Toyoda, pemmaraju}. 
The correct $E_{gap}$ of ZnO is also obtained with hybrid 
functionals (HY)~\cite{walsh, sarsari, patterson, Badaeva2008}. 
Linear response time-dependent density functional theory provides 
excited state energies, and it was recently 
applied to ZnO:Co~\cite{Badaeva2009, may}.  
However, the computational cost of those methods is much higher than that 
of LDA+$U$ and GGA+$U$. The relatively low cost of $+U$ methods is 
important in the context of high-throughput computations 
aimed at, e.g., optimization of selected material properties for 
applications.

\begin{table*}
\caption{\label{tabI}
Calculated ZnO band gap and energies of the $\textrm{Co}^{2+}$ levels. 
The symbol "$\sim$VBM" denotes the cases when the majority spin states 
are close to the VBM, or are degenerate with the valence bands. Several 
values are inferred from the figures showing density of states, and thus 
they are of limited accuracy. All values are in eV.}
\begin{ruledtabular}
\begin{tabular}{@{}l c c c c c l}
Ref. & method & $E_{gap}$ & $t_{2\uparrow}$ & $e_{2\downarrow}$ & $t_{2\downarrow}$ & 
comments\\
\hline
\cite{spaldin} 		& LDA & $<2$ & $\sim$VBM & 1.0 & 2.2\\
\cite{walsh} 		& GGA & 0.8 & $\sim$VBM & 0.8 & 2.0\\
\cite{Gilliland2012}& GGA & 1.2 & 0.3       & 1.3 & 3.0\\
\hline
\cite{Gopal} 	& LDA$+U$ & 0.8 & $\sim$VBM & 1.0 & 4.0 & {$U(\textrm{Zn})=0$, 
$U(\textrm{Co})=4.0$~eV}\\
\cite{Iusan} 	& LDA$+U$ & 2.0 & $\sim$VBM & 0.1 & 4.0 & 
{$U(\textrm{Zn})=9.0$~eV, $U(\textrm{Co})=5.0$~eV}\\
\cite{sarsari} 	& GGA$+U$ & 1.6 & $\sim$VBM & 1.2 & 3.3 & 
{$U(\textrm{Zn})=7.0$~eV, $U(\textrm{Co})=2.0$~eV}\\
\cite{Lany} & GGA$+U+\textrm{NLEP}$ & 3.3 & $\sim$VBM & 0.8 & 4.5 & 
{$U(\textrm{Zn})=2.8$~eV, $U(\textrm{Co})=7.0$~eV}\\
\hline
\cite{Toyoda} 	    & SIC & 3.0 & $\sim$VBM & 1.0 & 3.5\\
\cite{pemmaraju}	& SIC & 3.0 & 0.5       & 1.0 & 4.0\\
\hline
\cite{patterson} 	& HY & 3.3 & $\sim$VBM & 0.8 & 6.0\\
\cite{Badaeva2008}	& HY & 4.5 & $\sim$VBM & 0.5 & 6.0 & results for quantum dot $(\textrm{Zn}_{140}\textrm{Co})\textrm{O}_{141}$\\
\cite{walsh} 		& HY & 3.4 & $\sim$VBM & 1.5 & 5.0\\
\hline
\cite{sarsari} & HY+GW$_0$ & 3.3 & $\sim$VBM & 0.8 & 7.1\\
\hline
present & LDA$+U$ & 3.3 & 0.3 & 1.3 & 5.0 & {$U(\textrm{Zn})=12.5$~eV, 
$U(\textrm{O})=6.25$~eV,   $U(\textrm{Co})=3.0$~eV}\\
	    & LDA$+U$ & 3.3 & $\sim$VBM & 0.9 & 5.2 & {$U(\textrm{Zn})=12.5$~eV, 
$U(\textrm{O})=6.25$~eV,   $U(\textrm{Co})=4.0$~eV}\\
\end{tabular}
\end{ruledtabular}
\end{table*}
\normalsize

Table~\ref{tabI} summarizes the calculated values of $E_{gap}$ of ZnO 
and the energies of the three $\textrm{Co}^{2+}$ gap 
levels. They are qualitatively similar, predicting the majority-spin 
levels close to the valence 
band maximum (VBM), the minority $e_{2\downarrow}$ level in the lower 
half of the 
band gap, and the crystal field  splitted $t_{2\downarrow}$ level higher 
in energy. 
Quantitatively, however, the discrepancy between various methods exceeds 
4~eV, which is only 
partially explained by the large band gap error of the LDA and GGA. 
Indeed, 
even the corrected 
approaches, such as SIC or LDA+$U$, lead to differences larger by more 
than 1~eV. 

Correctness of a theoretical approach is assessed by comparing the 
results of calculations to experiment. This issue was not discussed 
in the quoted works, except Refs~\cite{Badaeva2008, Badaeva2009, may}.
In the case of ZnO:Co, experiment includes optical, transport, and 
magnetic measurements, and the corresponding observables are 
optical transition energies, thermal ionization energies, and 
magnetic moments and couplings.
Importantly, energies of excited states cannot be inferred from the 
differences in one-electron energies because of the strong intrashell 
coupling of $d(\textrm{Co})$ electrons, large lattice relaxations induced 
by the change of the Co charge state, and the presence of 
non-negligible (albeit small in ZnO) electron-hole coupling. A striking 
example is that of the internal $d(\textrm{Co})$ excitation: its 
experimental energy is 2.0~eV, while the 
$e_{2\downarrow}$-$t_{2\downarrow}$ energy difference obtained with 
hybrid functionals is more than twice higher, see Table~\ref{tabI}. 
This feature is 
also important when analyzing ionization (charge transfer) 
transitions, since the Co levels are sensitive to its charge state, 
and thus the calculations should be extended to charged Co. 
This was performed in Refs~\cite{Gluba, Raebiger} which calculated the 
transition levels. 

The aim of this paper is to provide a theoretical interpretation of the 
measured optical properties of ZnO:Co. To this end, we focus on the 
relevant transition energies rather than one electron levels. 
We also consider the change of the band gap of ZnO induced by Co.
The paper is organized as follows. In Section~\ref{sec2}, we present the 
method of calculations, 
including the calculations of the $U(\textrm{Co})$ term. 
The results of the electronic structure 
and optical transitions of ZnO:Co are shown in Section~\ref{sec3a} and 
\ref{sec3b}. The comparison of calculated results with experiment and 
their interpretation, given in Section~\ref{sec3c}, allow to find 
the optimal value of $U(\textrm{Co})$. 
Section~\ref{sec4} summarizes the obtained results.

\section{\label{sec2}Calculation details}
\begin{figure}[t!]
\begin{center}
\includegraphics[width=8.3cm]{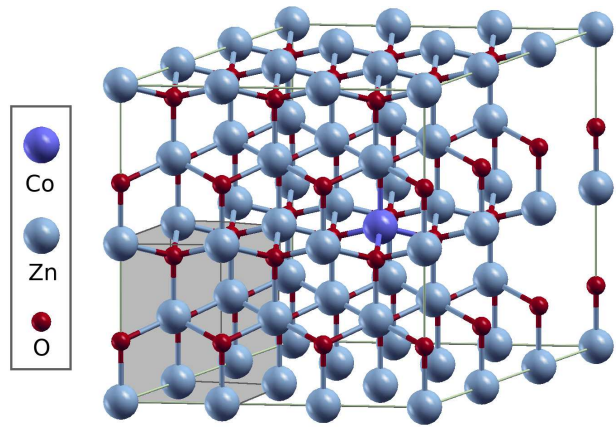}
\end{center}
\caption{\label{fig:Co-crystal} 
$3\times 3\times 2$ supercell of ZnO with Co impurity. 
Shaded region indicates a single unit cell of pure ZnO with 4 atoms.
}
\end{figure}

The calculations are performed within the density functional theory in 
the GGA approximation of the exchange-correlation 
potential~\cite{Hohenberg, KohnSham, PBE}, supplemented by the $+U$ 
corrections~\cite{Cococcioni, Anisimov1991, Anisimov1993}. We use the 
pseudopotential method implemented in the {\sc Quantum ESPRESSO} 
code~\cite{QE}, with the valence atomic configuration $3d^{10}4s^2$ for 
Zn, $2s^2p^4$ for O and $4s^2p^0 3d^7$ for Co. The plane-waves kinetic 
energy cutoffs of 30~Ry for wavefunctions and 180~Ry for charge density 
are employed. Spin-orbit coupling is neglected.

The electronic structure of the wurtzite ZnO is examined with a $8\times 
8\times 8$ $k$-point grid. Analysis of the Co impurity is performed using 
$3\times 3\times 2$ supercells with 72 atoms 
shown in Fig.~\ref{fig:Co-crystal}, and the $k$-space 
summations performed with a $3\times 3\times 3$ $k$-point grid. 
Larger supercells ($3\times 3\times 4$ with 144 atoms) 
and smaller supercells ($2\times 2\times 2$ with 32 atoms 
and $2\times 2\times 1$ with 16 atoms) are employed to obtain the dependence of 
the energy gap on the Co concentration. 
The ionic positions of ZnO:Co are optimized until the forces acting on 
ions became smaller than 0.02~eV/\AA. The calculated lattice constants of 
ZnO, $a= 3.23$~\AA\ and $c = 5.19$~\AA, as well as internal parameter 
$u = 0.38$, are underestimated by less than 1~\% in comparison with 
experimental values~\cite{Karzel, Ozgur}. The calculated average 
Co-O bond lengths for $\textrm{Co}^{2+}$ are 1.99 \AA, close to 
those of Zn-O, 1.97 \AA.

The underestimation of the band gap of ZnO is corrected by applying the 
$U$ term to $d(\textrm{Zn})$ and $p(\textrm{O})$ electrons. We find that 
the corrections $U(\textrm{Zn})=12.5$~eV and $U(\textrm{O})=6.25$~eV 
reproduce not only the experimental ZnO gap of 3.3~eV~\cite{Dong, Izaki, 
Srikant}, but also the width of 6~eV of the upper valence band of mostly 
$2p(\textrm{O})$ character, and the energy of the $d(\textrm{Zn})$ band, 
centered about 8~eV below the VBM~\cite{Lim}. Those values of the $U$ 
terms were tested by us for ZnO:Mn and ZnO:Fe~\cite{Mn, Fe}. 
$U(\textrm{O})$ directly opens the gap since the VBM is mainly derived 
from the $p(\textrm{O})$ orbitals, while $U(\textrm{Zn})$ changes the 
position of the  $d(\textrm{Zn})$-derived band well below the VBM. 
In previous works~\cite{Ma, Calzolari}, $U(\textrm{Zn})= 10-12$~eV and 
$U(\textrm{O})=6-7$~eV were proposed, while in Ref.~\cite{Agapito} 
$U(\textrm{Zn})= 12.8$~eV and $U(\textrm{O})=5.29$~eV were calculated by 
using pseudohybrid Hubbard density functional method. This consistency 
between the results of various approaches provides a complementary 
justification for our $+U$ values. 
Finally, we mention that 
by the band gap $E_{gap}$ we understand the 
single-particle band gap, which is equal to the energy difference between 
the Kohn-Sham energies of the conduction band minimum (CBM) and the 
VBM. 
As it was discussed in Ref.~\cite{Lany2008}, this also corresponds to 
the quasiparticle band gap. Inclusion of excitonic 
effects would be necessary in a detailed study of the optical response of 
ZnO, but this problem is outside the scope of this work.

The calculated total energies for all the considered charge 
states are used to obtain the transition levels 
$\varepsilon(q/q')$ between various charge 
states of Co. In the case of 
charged supercells, the image charge corrections and potential alignment 
are included in the calculations according to~\cite{Lany2008, Lany2009}.

The value of the $U$ correction for $d(\textrm{Co})$ is 
obtained in two ways. 
First, it is considered as a free parameter 
varying from 0 to 6~eV. The best agreement with the experimental optical 
transition energies is obtained for $U(\textrm{Co})=3$~eV.  
Second, we also compute $U(\textrm{Co})$ by linear response 
approach proposed in 
Ref.~\cite{Cococcioni}. We add small potential shifts that 
act only on the localized $d$ orbitals of Co through a projection 
operator, $\Delta V = \alpha \sum_{m} |\phi_m\rangle\langle\phi_m|$, and 
calculate variation of the Co occupations. The obtained (interacting and 
noninteracting) density response functions of the system with respect to 
these perturbations are used to compute the $U(\textrm{Co})$ term. 
In the above procedure, we use a supercell with 72 atoms and then 
extrapolate the results to larger supercell with over 300 Co ions. We 
obtain $U(\textrm{Co}) = 3.4$~eV, which is a little higher than the value 
fitted to experiment. 
Importantly, the constrained calculations with fixed initial 
occupations of $d(\textrm{Co})$ orbitals are needed, 
because final total energies depend on the 
initial fixed on-site occupation matrices. Therefore, we test all 
possible occupations of Co orbitals with integer occupation numbers 0 and 
1 to find the state with minimum energy for a given charge state $q$.

To find energies of optical transitions of Co in ZnO two approaches are 
used. The ionization energy for the transition $\textrm{Co}^{2+} \to 
\textrm{Co}^{3+}$ is obtained from the energy of $\varepsilon(+/0)$ level 
relative to the CBM, while that for $\textrm{Co}^{2+} \to 
\textrm{Co}^{1+}$  is obtained from the energy of $\varepsilon(0/-)$ 
relative to the VBM. 
A second approach is used to find energies of the 
excited configurations (obtained after e.g. the internal transition from 
$\textrm{Co}^{2+}$ to the excited ($\textrm{Co}^{2+})^*$ state). In this 
case, the occupations of the Kohn-Sham levels are fixed, and the 
Brillouin zone summations are approximated by the $\Gamma$ point 
values~\cite{QE}. 
Moreover, supercells remain electrically 
neutral even when the ionized $\textrm{Co}^{3+}$ is analyzed 
because of the presence of the excited electron in the conduction band.
In consequence, the electric fields generated by $\textrm{Co}^{3+}$ are 
largely screened, their impact on both excitation energies and the Co 
gap levels~\cite{Komsa} is expected to be small, 
and total energies need not to be corrected for the 
spurious defect-defect coupling. 
These two approaches give ionization energies 
that agree to within less than 0.1~eV.

\section{\label{sec3}Results}

\subsection{\label{sec3a}Co levels in ZnO}
\begin{figure}[t!]
\begin{center}
\includegraphics[width=8.3cm]{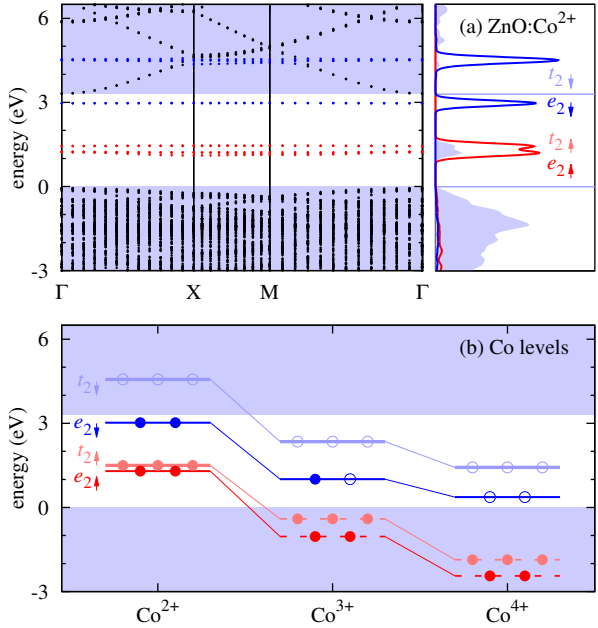}
\end{center}
\caption{\label{fig:Co-bands} 
(a) Band structure (left panel) and DOS (right panel) of 
ZnO:$\textrm{Co}^{2+}$. Zero energy is at the VBM. Gray areas display the 
total DOS, while red and blue lines indicate the DOS projected on 
$d_{\uparrow}(\textrm{C}o)$ and  $d_{\downarrow}(\textrm{Co})$ orbitals, 
respectively. 
(b) Kohn-Sham levels of Co in various charge states. 
Horizontal (gray) lines denote the band gap of ZnO. Electrons and holes 
on the $d(\textrm{Co})$ levels are denoted by filled and empty dots, 
respectively. Spin-up states of $\textrm{Co}^{3+}$ and $\textrm{Co}^{4+}$ 
are smeared out over the valence bands. 
The calculations are performed for $U(\textrm{Co})=0$. 
}
\end{figure}

Figure~\ref{fig:Co-bands}(a) shows the band structure and the density of 
states (DOS) of ZnO doped with $\textrm{Co}^{2+}$ for $U(\textrm{Co})=0$. 
$\textrm{Co}^{2+}$ introduces three levels in the gap: close-lying 
$e_{2\uparrow}$ doublet and $t_{2\uparrow}$ triplet at about 1.5~eV above 
the VBM, and the $e_{2\downarrow}$ doublet at 3.2~eV. The empty triplet 
$t_{2\downarrow}$ is degenerate with the conduction band. Actually, both 
$t_{2\uparrow}$ and $t_{2\downarrow}$ triplets are weakly split by about 
0.1~eV by the wurtzite crystal field, but we omit this effect for the 
sake of clarity. 
The strong dependence of the Kohn-Sham levels of Co on the 
charge state is clearly visible in Fig.~\ref{fig:Co-bands}(b). The levels 
of $\textrm{Co}^{2+}$ with seven $d$ electrons are 1-2~eV higher in energy 
than those of $\textrm{Co}^{3+}$ with 6 electrons, because the intrashell 
Coulomb repulsion increases with the increasing $d$-shell 
occupation~\cite{Mn, Fe, OptMat}. In particular, the spin-up states of 
$\textrm{Co}^{3+}$ are below the VBM, while both spin-down levels are in 
the gap: $e_{2\downarrow}$ at 1.1~eV and $t_{2\downarrow}$ at 2.4~eV 
above the VBM. The second ionization to $\textrm{Co}^{4+}$ farther 
decreases the Co levels.

\begin{figure}[t!]
\begin{center}
\includegraphics[width=8.3cm]{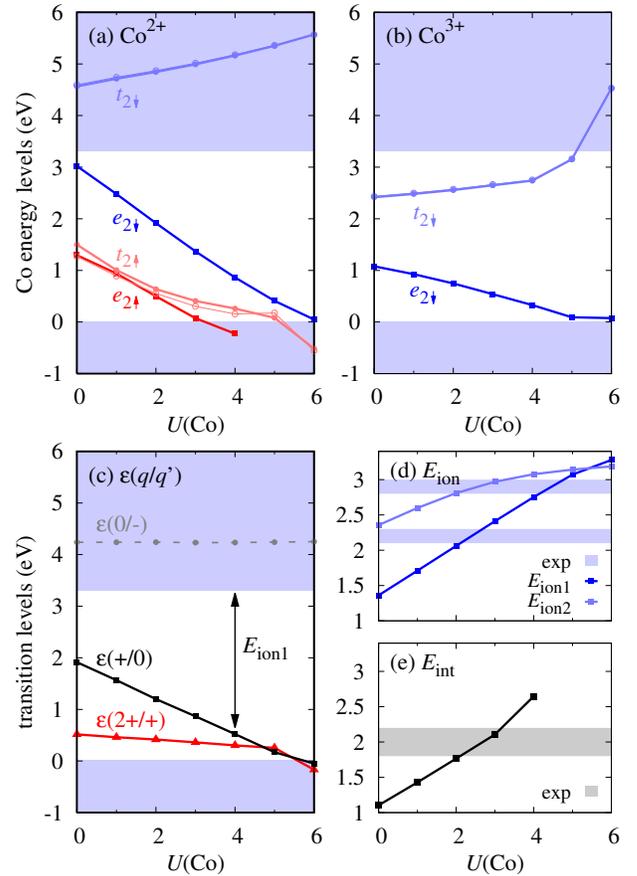}
\end{center}
\caption{\label{fig:Co-U} 
The Kohn-Sham energy levels of (a) $\textrm{Co}^{2+}$, (b) 
$\textrm{Co}^{3+}$ and (c) the transition levels $\varepsilon(q/q')$ 
calculated as a function of $U(\textrm{Co})$. The $\varepsilon(0/-)$ is 
only estimated, see in text for details. The $U(\textrm{Co})$ dependence 
of (d) ionization and (e) internal transition energies. 
Experimental results are shown in both cases: thresholds of ionization 
transitions are taken from photocurrent measurements~\cite{Kittilstved, Liu2005, 
Gamelin2010, Gamelin2011}, while the energy range of 
internal transitions is taken from absorption~\cite{Koidl, 
KIM2004, Singh2008, Jin, Matsui2013, Xu2013, Lecuna2014}.
See text for definitions of ionization $E_{ion1}$ and 
$E_{ion2}$ and internal transition $E_{int}$  energies.
}
\end{figure}

The dependence of $\textrm{Co}^{2+}$ and $\textrm{Co}^{3+}$ levels on 
$U(\textrm{Co})$ is presented in Fig.~\ref{fig:Co-U}(a) and 
\ref{fig:Co-U}(b), respectively. 
The $U$-induced contribution $V_U$ to the Kohn-Sham 
potential is~\cite{Cococcioni}
\begin{equation}
V_U|\psi_{k\nu}^\sigma\rangle = U \sum_{m,\sigma} 
(1/2 - \lambda_m^\sigma)|\phi_m\rangle\langle\phi_m|\psi_{k\nu}^\sigma\rangle,
\label{eq1} 
\end{equation}
where $\phi_m$ are the localized $d$ orbitals occupied by $\lambda_m$ 
electrons, and $\psi_{k\nu}^\sigma$ are the Kohn-Sham states for the 
wavevector $k$, band $\nu$, and spin $\sigma$. 
The $V_U$ potential only acts on the contribution of the $m$th 
$d$(Co) orbital to the given $(\nu,k,\sigma)$ state, 
and this contribution is evaluated by the appropriate projection 
according to Eq.~\ref{eq1}. 
Equation~\ref{eq1} also shows that there is a 
negative shift of the fully occupied $e_{2\uparrow}$, $t_{2\uparrow}$ 
and $e_{2\downarrow}$ levels of $\textrm{Co}^{2+}$, while the shift is 
positive for empty $t_{2\downarrow}$. 
For $U>4$~eV, the spin-up $\textrm{Co}^{2+}$ levels merge with the 
valence band. The energy levels of $\textrm{Co}^{1+}$ are not shown 
because this charge 
state is not stable and it will not be assumed by Co for realistic Fermi 
energies. Indeed, the $t_{2 \downarrow}$ of $\textrm{Co}^{2+}$ is at 
least 1~eV above the CBM, see Fig.~\ref{fig:Co-U}(a). 
We note that the exact energy of $t_{2\downarrow}$ relative to the CBM is 
of importance for the magnetic coupling between Co ions mediated by free 
carriers, see the discussion in Ref.~\cite{sanvito-comm}. 

The calculated $\textrm{Co}^{2+}$ energy levels 
are compared with the previous 
calculations in Table~\ref{tabI}. 
The LDA and GGA approximations lead to the underestimated $E_{gap}$ and, 
in consequence, to the $e_{2 \downarrow}$ and $t_{2 \downarrow}$ levels 
degenerate with the conduction band~\cite{spaldin, walsh, Gilliland2012, Gopal}. 
The $U(\textrm{Co})$ term shifts down all the occupied Co levels, 
and thus the $e_{2 \downarrow}$  
is in the band gap~\cite{Iusan, sarsari}, as in our case. 
However, ionization energies are too low since the gap is strongly 
underestimated even if $U(\textrm{Zn})$ is 
applied ~\cite{Iusan, sarsari}. 
On the other hand, our results obtained with $U(\textrm{Co}) = 3-4$~eV 
are in a reasonable agreement with bandgap corrected methods 
employed in Refs~\cite{ walsh, sarsari, Lany, Toyoda, pemmaraju,  patterson, Badaeva2008}, and in particular with the HY calculations.

The dependence of the Co levels on $U(\textrm{Co})$ is reflected in the 
corresponding dependence of the thermodynamic transition levels 
shown in Fig.~\ref{fig:Co-U}(c). 
The value of 
$\varepsilon(0/-)$ given in Fig.~\ref{fig:Co-U}(c) is only estimated, 
since the $t_{2 \downarrow}$ electron of $\textrm{Co}^{1+}$ would 
autoionize to the CBM. 
In turn, the $\varepsilon(+/0)$ and $\varepsilon(2+/+)$ levels are in the 
gap, indicating that $\textrm{Co}^{2+}$, $\textrm{Co}^{3+}$ and 
$\textrm{Co}^{4+}$ are possible stable charge states of Co in ZnO. 
Transition levels 
$\varepsilon(+/0)=0.85$~eV and 
$\varepsilon(0/-)= 4.5$~eV (i.e., 1.1~eV above the CBM) 
were obtained in Ref~\cite{Gluba}, and 
$\varepsilon(+/0)=0.4$~eV and 
$\varepsilon(0/-)= 3.7$~eV (i.e., 0.9~eV above the CBM)
in Ref~\cite{Raebiger}.  
Our results for $U(\textrm{Co})= 3-4$~eV are between those of 
Ref.~\cite{Gluba} and Ref. \cite{Raebiger}. 
The $\varepsilon(2+/+)$ transition level, which was not considered in
Refs~\cite{Gluba, Raebiger}, is practically degenerate with the VBM.  

\subsection{\label{sec3b}Optical transitions}
\begin{figure}[t!]
\begin{center}
\includegraphics[width=8.3cm]{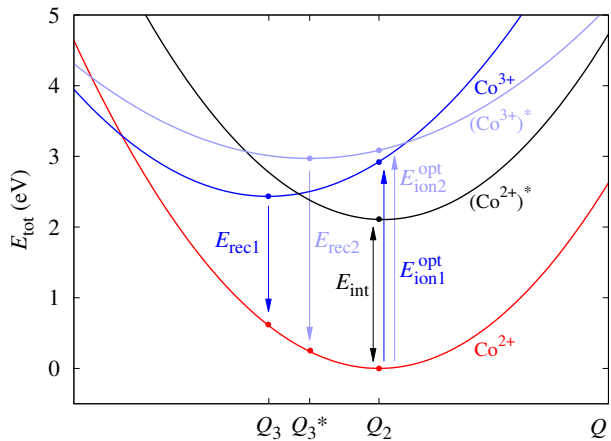}
\end{center}
\caption{\label{fig:Co-ene1} 
Total energy change of the $\textrm{Co}^{2+}$ and of the 
excited states of Co as a function of configuration coordinate $Q$. 
$\textrm{Co}^{3+}$ means
$(\textrm{Co}^{3+}, e_{CB})$, while $(\textrm{Co}^{3+})^*$ means 
$((\textrm{Co}^{3+})^*, e_{CB})$ state. $Q_2$, $Q_3$ and $Q_3^*$ are 
equilibrium atomic configurations of $\textrm{Co}^{2+}$, 
$\textrm{Co}^{3+}$ and $(\textrm{Co}^{3+})^*$ charge states, associated 
with Co-O bond lengths. $U(\textrm{Co})=3$~eV is assumed.
Ionization and recombination energies are defined in the text. 
}
\end{figure}

\begin{figure}[t!]
\begin{center}
\includegraphics[width=8.3cm]{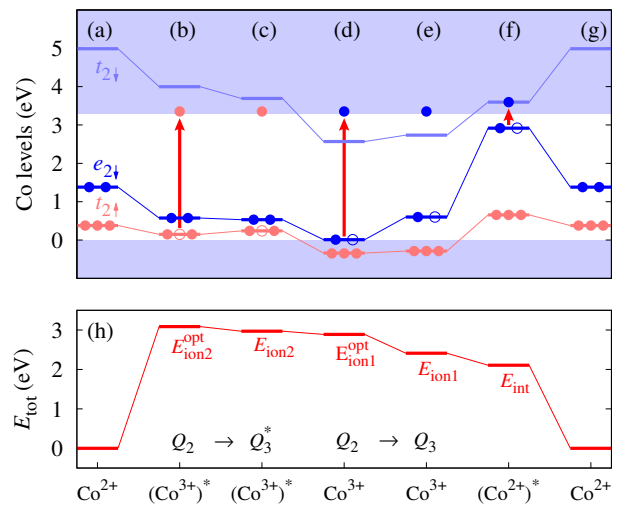}
\end{center}
\caption{\label{fig:Co-ene2} 
(a-g) The changes of $t_{2\uparrow}$, $e_{2\downarrow}$ and 
$t_{2\downarrow}$ levels induced by the ionization and intrashell 
transitions. 
The levels of Co at $Q_2$ correspond to vertical 
transitions, while these ones at $Q_3$ and $Q_3^*$ include relaxation of 
$\textrm{Co}^{3+}$ and $(\textrm{Co}^{3+})^*$ to their equilibrium 
configurations. 
Total energies relative to that of $\textrm{Co}^{2+}$ are shown in panel 
(h). $U(\textrm{Co})=3$~eV is assumed. 
}
\end{figure}

We consider four optical excitation processes, namely three ionization 
channels and the internal excitation, together with the corresponding 
recombination transitions. 
We also study the composition dependence of the $E_{gap}$. 
Figure~\ref{fig:Co-ene1} shows the 
configuration diagram, where the configuration coordinate $Q$ should be 
regarded as the average Co-O bond length for a given electronic 
configuration of Co. 
The involved systems are $\textrm{Co}^{2+}$, its excited state 
$(\textrm{Co}^{2+})^*$, as well as $\textrm{Co}^{3+}$ and its excited 
$(\textrm{Co}^{3+})^*$ state. 
The corresponding equilibrium configuration coordinates are $Q_2$, 
$Q_2^*$,
 $Q_3$, and $Q_3^*$. 
$U(\textrm{Co})=3$~eV is assumed.

1. The first ionization channel consists in the vertical optical 
ionization of $\textrm{Co}^{2+}$ with the energy $E^{opt}_{ion1}$, when 
an electron is transferred from the $e_{2\downarrow}$ level to the CBM,  
\begin{equation}
\textrm{Co}^{2+} + E^{opt}_{ion1}   \to (\textrm{Co}^{3+},e_{CB}). 
\label{ion1} 
\end{equation}

\noindent
During calculations for this process, the occupation of the 
initial state is fixed to be 
$(e_{2\uparrow}^2t_{2\uparrow}^3e_{2\downarrow}^2t_{2\downarrow}^0)$, 
while that of the final state is 
$(e_{2\uparrow}^2t_{2\uparrow}^3e_{2\downarrow}^1t_{2\downarrow}^0,e_{CB}
)$, 
see also Fig.~\ref{fig:Co-ene2}(a), Fig.~\ref{fig:Co-ene2}(d-e), 
and Fig. 4. 
The calculated optical ionization energy $E_{ion1}^{opt}$ 
of the vertical transition is 2.9~eV.
After the excitation, the system can 
relax from the equilibrium configuration $Q_2$ of $\textrm{Co}^{2+}$ 
toward that of $\textrm{Co}^{3+}$, $Q_3$, which lowers the total energy 
by 0.5~eV. 
Thus, the onset of the absorption is predicted 
to occur at the zero phonon line energy $E_{ion1}=$2.4~eV. 

The change of the charge state causes the reduction of Co-O bonds by 
about 5 per cent, which is reflected in the difference between $Q_2$ and 
$Q_3$ in Fig.~\ref{fig:Co-ene1}. 
The photoelectron recombines 
[$(\textrm{Co}^{3+}, e_{CB})\to \textrm{Co}^{2+}$] with the recombination energy $E_{rec1}$. 
The energy $E_{rec1}$ of $\textrm{Co}^{3+}$ at equilibrium is 1.8~eV. 
After recombination, the system relaxes to $Q_2$ releasing 0.6~eV.

Experimentally, luminescence energies are affected by the ratio of the 
radiative recombination rate to the lattice relaxation time of phonon 
emission. When the recombination is fast, the system does not relax to 
equilibrium, and the absorption and emission energies are almost equal to 
each other. In the opposite limit, long recombination times allow for the 
full relaxation of atomic configurations by phonon emission, which 
increases the difference between $E_{ion1}^{opt}$ and $E_{rec1}$, i.e., 
the Franck-Condon shift.

2. The second ionization channel consists in the ionization of 
$\textrm{Co}^{2+}$ via electron transfer from $t_{2\uparrow}$ to 
the CBM: 
\begin{equation}
\textrm{Co}^{2+} + E^{opt}_{ion2}   \to  ((\textrm{Co}^{3+})^*, e_{CB}). 
\label{ion2} 
\end{equation}

\noindent
The occupations in the excited state $(\textrm{Co}^{3+})^*$ are 
$(e_{2\uparrow}^2t_{2\uparrow}^2e_{2\downarrow}^2t_{2\downarrow}^0, e_{CB})$, 
as indicated in Fig.~\ref{fig:Co-ene2}(b-c). 
The corresponding ionization and recombination energies are 
$E^{opt}_{ion2}$ and $E_{rec2}$. 
Ionization of $\textrm{Co}^{2+}$ to  
$(\textrm{Co}^{3+})^*$ produces a $\sim 3$ per cent shortening of Co-O 
bonds,  which is smaller than in the case of $\textrm{Co}^{3+}$. 
For this 
reason, the relaxation energy is smaller, about 0.1~eV, and the vertical 
and zero phonon absorption energies are similar, 
$E^{opt}_{ion2}= 3.1$~eV and $E_{ion2}= 3.0$~eV, 
respectively. The Franck-Condon shift between absorption and emission 
lines is also smaller since the recombination energy at the 
configuration $Q_3^*$ is $E_{rec2}= 2.7$~eV.

3. The third possible ionization channel consists in the ionization of 
$\textrm{Co}^{2+}$ when an electron is transferred from the VBM to 
$t_{2\downarrow}$ leaving a hole in the valence band, $h_{VB}$:

\begin{equation}
\textrm{Co}^{2+} + E^{opt}_{ion3}   \to  (\textrm{Co}^{1+}, h_{VB}). 
\label{ion3} 
\end{equation}

\noindent
In that case, fixed occupations for ionized state are 
$(h_{VB},e_{2\uparrow}^2t_{2\uparrow}^3e_{2\downarrow}^2t_{2\downarrow}^1)$.
As it was pointed out above and shown in Fig.~\ref{fig:Co-U}(c), 
the transition level $\varepsilon(0/-)$ lies 
above the CBM, and thus the 
$\textrm{Co}^{1+}$ charge state is not stable. The photoionization energy 
$E_{ion3}$ is above 4~eV, higher than $E_{gap}$.

4. The internal excitation of $\textrm{Co}^{2+}$, 
\begin{equation}
\quad\textrm{Co}^{2+} +  E_{int}\to (\textrm{Co}^{2+})^*,
\label{INT} 
\end{equation}

\noindent
consists in the transfer of an electron from the doubly occupied 
$e_{2\downarrow}$ to the empty $t_{2\downarrow}$ level, thus 
$(e_{2\uparrow}^2t_{2\uparrow}^3e_{2\downarrow}^1t_{2\downarrow}^1)$ is 
fixed for $(\textrm{Co}^{2+})^*$ as shown in Fig.~\ref{fig:Co-ene2}(f). 
The 
corresponding excitation energy is denoted by $E_{int}$, and we find 
$E_{int}=2.1$~eV. In this case, $Q_2  \approx Q_2^*$,
since the charge state of the dopant remains unchanged and the 
redistribution of $d(\textrm{Co})$ 
electrons affects the bonds by less than 0.1\%.

A further insight into those processes can be gained from 
Fig.~\ref{fig:Co-ene2}, which shows the $t_{2\uparrow}$, 
$e_{2\downarrow}$ and $t_{2\downarrow}$ levels calculated for six cases: 
$\textrm{Co}^{2+}$ at equilibrium $Q_2$ (panels (a) and 
(g)), the photoionized $((\textrm{Co}^{3+})^*, e_{CB})$ at the $Q_2$ and 
$Q_3^*$ configurations (panels (b) and (c)), the $(\textrm{Co}^{3+}, 
e_{CB})$ at $Q_2$ and $Q_3$ (panels (d) and (e)), and finally the 
$(\textrm{Co}^{2+})^*$ state at $Q_2$ (panel (f)). The excited states are 
ordered from the highest to the lowest total energies, which are given in 
panel (h). 
The dependence of the Kohn-Sham levels of Co on the charge 
state follows from two effects. The first and the dominant one is the 
reduced Coulomb intrashell repulsion characterizing the ionized 
$\textrm{Co}^{3+}$ and $(\textrm{Co}^{3+})^*$, which induces downward 
shift of gap levels relative to those of the neutral 
$\textrm{Co}^{2+}$. 
The second effect is the upward shift of the energies of Co levels, 
which are induced by the decrease of Co-O bond lengths from 
$Q_2$ to $Q_3$ or $Q_3^*$, see also Ref.~\cite{Mn}. 
Both effects are stronger for $\textrm{Co}^{3+}$, where 
the optical ionization decreases the position of Co levels by 1.5~eV, 
while the relaxation from $Q_2$ to $Q_3$ rises the Co levels by $\sim 
0.7$~eV. 
The changes induced by the internal transition, shown in the panels 
(g)-(f), have a different origin. In this case, the energy shifts of 
$e_{2\downarrow}$ and $t_{2\downarrow}$ are comparable, and are caused 
mainly by the $U(\textrm{Co})$ correction. They have opposite signs 
because, in agreement with Eq.~\ref{eq1}, after the excitation from 
$\textrm{Co}^{2+}$ to $(\textrm{Co}^{2+})^*$, the occupation of 
$t_{2\downarrow}$ increases by 1 and that of $e_{2\downarrow}$ decreases 
by 1.

Considering recombination processes we see that after the internal 
excitation of $\textrm{Co}^{2+}$, the $t_{2\downarrow}$ occupied with 
one electron is about 0.25~eV above the CBM. This suggests that the 
excited 
$(\textrm{Co}^{2+})^*$ can spontaneously ionize, releasing one electron 
to the CBM in the reaction [$(\textrm{Co}^{2+})^*\to (\textrm{Co}^{3+}, 
e_{CB})$], i.e., a transition from (f) to (e) should spontaneously occur. 
However, the total energy of the latter state is 0.3~eV $higher$ than the 
energy of $(\textrm{Co}^{2+})^*$, see Fig.~\ref{fig:Co-ene2}(h), and  
therefore the ionization is possible, but it is a thermally activated 
process. 
Second, 
the results of Fig.~\ref{fig:Co-ene1} show that there are two possible 
channels of recombination for $\textrm{Co}^{3+}$ and 
$(\textrm{Co}^{3+})^*$. The first one is the one-step direct 
recombination with $E_{rec1}$ or $E_{rec2}$, while the second channel is 
a two-step process, in which the electron capture on $t_{2\downarrow}$ or 
$e_{2\uparrow}$ is followed by the internal deexcitation.

The above results illustrate the fact that excitation energies cannot be 
estimated based on the single-electron energies. 
In particular, 
the calculated internal transition energy is 2.1~eV, while the 
energy difference between the $\textrm{Co}^{2+}$ levels is almost twice 
larger, 3.5~eV. 
Such a large discrepancy is expected to hold also for hybrid 
functionals calculations~\cite{walsh, sarsari, patterson, Badaeva2008}.

Comparing our results with the recent study in Ref.~\cite{may} 
using 
the linear response time dependent density functional theory, we note 
that 
their results for the excitation to $(\textrm{Co}^{2+})^*$ and  
$\textrm{Co}^{3+}$ states were obtained for small quantum dots and 
extrapolated to bulk ZnO, nevertheless they agree with our values to 
within 0.2~eV.

\begin{figure}[t!]
\begin{center}
\includegraphics[width=8.3cm]{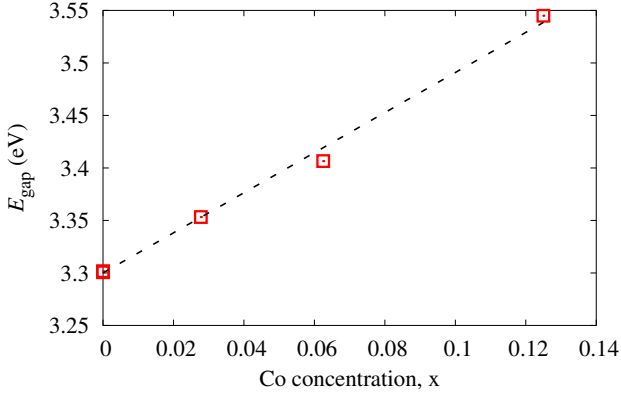}
\end{center}
\caption{\label{fig:Co-Egap} 
The Kohn-Sham band gap as a function of Co concentration in 
Zn$_{1-x}$Co$_x$O. 
$U(\textrm{Co})=3$~eV is assumed. Squares denote suprecell 
results, and the dashed line shows the fitted dependence 
$E_{gap}(x)=E_{gap}(0)+bx$, with $E_{gap}(0)=3.3$~eV and $b=1.9$~eV. 
}
\end{figure}

5. Finally, the observed optical properties of ZnO:Co 
include also fundamental excitation and recombination transitions
with energy given by the band gap. 
As it is shown in Fig.~\ref{fig:Co-Egap}, 
the calculated $E_{gap}$ increases with the increasing Co concentration.  
The detailed analysis of those transitions must include formation 
of excitons (which binding energy in ZnO is 60~meV~\cite{Reynolds}), but this issue 
is out of the scope of this work.

\subsection{\label{sec3c}Comparison with experiment}

The comparison of our results with the experimental data is satisfactory. 
The calculated dependencies of $E_{ion}$  and $E_{int}$ on 
$U(\textrm{Co})$ allow to find its optimal value. The results are 
shown in Fig.~\ref{fig:Co-U}(d) and (e). A good agreement with experiment 
is obtained with $U(\textrm{Co})=3.0$~eV. In particular:

(i) The observed increase of the band gap with the Co concentration $x$ 
is well reproduced, see Fig.~\ref{fig:Co-Egap}. The calculated 
coefficient $b$ defined by the relation $E_{gap}(x) = E_{gap}(0) + bx$ is 
1.9~eV, which agrees well with the experimental values 
1.1~\cite{KIM2004}, 1.7~\cite{Pacuski:PRB2006}, 2.3~\cite{Gamelin2011} 
and 2.5~eV~\cite{Matsui2013}.

(ii) Our energy of the internal transition $E_{int}=2.1$~eV is close to 
the experimental $\sim 2.0$~eV, which is seen in both absorption and 
luminescence. The observed splitting of this line is partly due to the 
crystal field splitting of the $t_{2\downarrow}$ level, and partly due to 
the spin-orbit coupling, which is neglected in our 
calculations~\cite{footnote}.

(iii) The optical ionization of $\textrm{Co}^{2+}$  is observed in a 
broad band which begins at about 2.2~eV, extends up to about 3~eV, and 
was monitored in photocurrent by Gamelin {\it et al.}~\cite{Kittilstved, Liu2005, 
Gamelin2010, Gamelin2011}. Their further analysis revealed 
that it originates in two transitions, the lower energy one assigned to 
[$\textrm{Co}^{2+}\to (\textrm{Co}^{3+},e_{CB})$], and the higher energy 
transition interpreted as [$\textrm{Co}^{2+}\to (\textrm{Co}^{1+},h_{VB})$]. 

\begin{figure}[t!]
\begin{center}
\includegraphics[width=8.3cm]{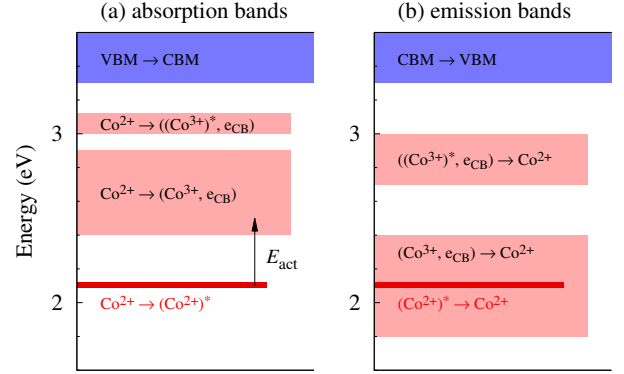}
\end{center}
\caption{\label{fig:Co-AbsEmis} 
Calculated (a) absorption and (b) emission bands. 
}
\end{figure}

According to our results, the ionization energy corresponding to the 
zero-phonon transition [$\textrm{Co}^{2+}\to (\textrm{Co}^{3+},e_{CB})$] 
is about 2.4~eV, and the phonon-assisted transitions extend 
up to $E_{ion1}^{opt}=2.9$~eV. 
Both values are higher by about 0.2~eV relative to experiment. This 
reasonable agreement confirms the assignment proposed in 
Refs~\cite{Kittilstved, Liu2005, Gamelin2010, Gamelin2011}. 
This 2.4-2.9~eV absorption band is shown in Fig.~\ref{fig:Co-AbsEmis}(a). 
Its width can also be inferred from Fig.~\ref{fig:Co-ene1}. 

On the other hand, the obtained results do not support 
the identification of the 
second ionization transition as [$(\textrm{Co}^{2+})\to (\textrm{Co}^{+}, 
h_{VB})$], in which an electron is excited from the VBM to 
$\textrm{Co}^{2+}$ level at energies above 2.7~eV~\cite{Kittilstved, Liu2005, Gamelin2010, Gamelin2011}. 
Indeed, our results show that this transition requires at least 4~eV, 
as it follows from the energy of the $\varepsilon(0/-)$ transition 
level relative to the VBM (see Fig.~\ref{fig:Co-U}(c)). 
Instead, we propose that the observed higher absorption band 
originates in  [$(\textrm{Co}^{2+})\to ((\textrm{Co}^{+3})^*, e_{CB})$] 
transition from the Co spin-up state with the ionization energy in the 
range 3.0-3.1~eV shown in Fig.~\ref{fig:Co-AbsEmis}(a). 

In the absorption measurements, only one ionization channel is seen at 
energies just below the band gap~\cite{Tuan2004, Pacuski:PRB2006, Gilliland, 
Matsui2013, Guo2015}. Based on our results, we assign this 
transition to $t_{2 \uparrow} \to$ CBM, $i.e.$ to  [$\textrm{Co}^{2+}\to 
(\textrm{Co}^{3+})^*,e_{CB})$], because its energy fits the experiment. 
Moreover, the second possible transition, $e_{2\downarrow} \to$ CBM, 
is expected to be forbidden.

(iv) According to our results, after the internal excitation 
$(\textrm{Co}^{2+})^*$ can ionize releasing an electron to the CBM, and 
the corresponding activation energy is 
$E_{act}=0.3$~eV.  The experiments of 
Refs~\cite{Liu2005, Gamelin2010, Gamelin2011}  revealed that 
$(\textrm{Co}^{2+})^*$ indeed ionizes, because the internal  transition 
results in photoconductivity. The observed ionization has a thermally 
activated character with energy of about 50~meV, which is 
somewhat smaller than the calculated value. 

The calculated absorption and emission bands are summarized in Fig.~\ref{fig:Co-AbsEmis}. 
The lowest energy excitation at 2.1~eV is associated with the internal transition to excited $\textrm{Co}^{2+}$. Next, the two sub-bands corresponding the ionization transitions from $\textrm{Co}^{2+}$ to $\textrm{Co}^{3+}$ and $(\textrm{Co}^{3+})^*$ appear at 2.4-2.9~eV and at 3.0-3.1~eV. 
Each subband extends from the zero-phonon transition at energy $E_{ion 1(2)}$ (which includes atomic relaxations around Co) to the line at $E^{opt}_{ion 1(2)}$ of the vertical transition ($i.e.$, without changes in atomic positions). 
At the highest energies, above 3.3~eV, the $\textrm{VBM}\to\textrm{CBM}$ transitions occurs. The energy range of charge transfer and ZnO band-to-band transitions are expected to change with the Co concentration. 
In the emission bands, the transitions from $\textrm{Co}^{3+}$ and $(\textrm{Co}^{3+})^*$ to $\textrm{Co}^{2+}$ occur at 1.8-2.4~eV and 2.7-3.0~eV, respectively. 
Here again, the finite width of subbands spans the energy window between 
the zero-phonon lines $E_{ion 1(2)}$ and the vertical transitions $E_{rec 1(2)}$. 
(The widths of both absorption and emission bands can be inferred from Fig.~\ref{fig:Co-ene1}, 
where the vertical transitions are shown by arrows.) 
Both transitions can be completely suppressed by the two step processes involving $(\textrm{Co}^{2+})^*$, which is probably observed in the luminescence~\cite{Schulz, Matsui2013, Xu2013, Lecuna2014, Lommens}. 

\section{\label{sec4}Conclusions}
The results of the GGA$+U$ calculations 
explain the available experimental properties of substitutional Co in 
ZnO. 
In particular, the calculated composition dependence of the ZnO:Co band 
gap agrees well with the experimental data~\cite{KIM2004,Pacuski:PRB2006,Matsui2013,Gamelin2011}.
While the n-doping of ZnO:Co does not change the 
charge state of Co, the p-doping will be compensated because Co can act 
as a double donor.

We considered four possible optical transitions involving 
$\textrm{Co}^{2+}$. The first one is the internal $d-d$ transition, which  
calculated transition energy, 2.1~eV, corresponds well with the 
experimental value 2.0~eV. Next, there are two ionization transitions 
$\textrm{Co}^{2+} \to \textrm{Co}^{3+}$, in which an electron is 
transferred from one of the two $d(\textrm{Co})$ gap states to the 
conduction band. The lower energy channel is related with the excitation 
from $e_{2\downarrow}$ to CBM, with energies in the range 2.9-2.4~eV, in 
which the upper limit corresponds the zero phonon line. The higher energy 
channel is related with the transition $t_{2\uparrow} \to \textrm{CBM}$ 
with energies 3.1-3.0~eV, and leaving Co in the excited 
state $(\textrm{Co}^{3+})^*$. Both excitation channels occur in 
parallel, and extent from 3.1~eV to 2.4~eV. This agrees to within 0.2~eV 
with observations. The fourth possible ionization process consists in 
exciting an electron from the valence band to the $t_{2\downarrow}$ Co 
state, [$(\textrm{Co}^{2+})\to (\textrm{Co}^{+}, h_{VB})$]. This process 
was suggested in Refs~\cite{Kittilstved, Gamelin2010, Gamelin2011}. 
According to our results, the corresponding ionization energies are 
higher than 4~eV, which questions this interpretation.

We also point out that there are two recombination channels of 
photoelectrons, the direct recombination, and a two-step process in which 
the photoelectron is first captured by $\textrm{Co}^{3+}$, and then 
recombines via the internal $d-d$ transition.

The excitation-recombination processes are strongly affected by the 
intrashell Coulomb coupling. Manifestation of the coupling is provided by 
the pronounced dependence of the Co levels on their occupations. The 
levels of ionized $\textrm{Co}^{3+}$  and $(\textrm{Co}^{3+})^*$ are 
lower than those of $\textrm{Co}^{2+}$ due to the weaker intrashell 
Coulomb repulsion for an ion with smaller number of $d$-electrons. 
Additionally, both charge transfer transitions involve large lattice 
relaxations, which also influence the dopant levels. 
On the other hand, in spite of the fact the internal transition does not 
change the charge state, the $\textrm{Co}^{2+}$ levels are shifted by 
about 1~eV. In this case, the effect has a different origin, namely the 
occupation-dependent $U(\textrm{Co})$ corrections.

The theoretical level energies of Co depend on one parameter, $U(\textrm{Co})$. Its value is 
established in two ways. First, we treat $U$ as a fitting parameter. This allows us to reproduce the 
two measured ionization energies and the energy of the internal transition to within 0.1-0.2~eV 
with a single value $U(\textrm{Co})=3.0$~eV. Moreover, the discussed dependence of 
transition energies on $U$ provides an additional insight into the impact of the $+U$ 
corrections on the electronic structure of transition metal ions in semiconductors. In the second 
approach, the theoretical value $U(\textrm{Co})=3.4$~eV is obtained by the linear  response 
method of Ref.~\cite{Cococcioni}. It leads to somewhat less accurate energies of the optical 
transitions than the optimal fitted value 3.0~eV, but the agreement between the two methods is 
satisfactory. This farther confirms our identification of the observed transitions. Those results 
demonstrate that GGA+$U$ is an alternative to the linear response time dependent density 
functional theory of Ref.~\cite{may}.

\section*{Acknowledgements}
The authors acknowledge the support from the Project No. 
2016/21/D/ST3/03385, which are financed by Polish National Science 
Centre. Calculations were performed on ICM supercomputers of University 
of Warsaw (Grant No. G46-13 and G16-11).

\vspace*{0.5cm}


\providecommand{\newblock}{}

\end{document}